\pdfoutput=1
\documentclass[twocolumn,showpacs,preprintnumbers,amsmath,amssymb]{revtex4} 
\usepackage{graphicx}
\usepackage{color}
\usepackage[all]{xy}
 
\def\setR{\mathbb{R}}

\def\ie {i.e.} 

\newcommand{\norm}[1]{\parallel\!#1\!\parallel}

\newcommand{\dron}[3]{\frac{\partial^{#1} {#2}}{\partial{#3}^{#1}}}

\newcommand{\sss}[1]{\scriptscriptstyle #1}

\begin{document}

\title{Conformally invariant formalism for the electromagnetic field with currents in Robertson-Walker spaces}

\author{E. Huguet$^1$}
\author{J. Renaud$^2$}
\affiliation{$1$ - Universit\'e Paris Diderot-Paris 7, APC-Astroparticule et Cosmologie (UMR-CNRS 7164), 
Batiment Condorcet, 10 rue Alice Domon et L\'eonie Duquet, F-75205 Paris Cedex 13, France.  \\
$2$ - Universit\'e Paris-Est, APC-Astroparticule et Cosmologie (UMR-CNRS 7164), 
Batiment Condorcet, 10 rue Alice Domon et L\'eonie Duquet, F-75205 Paris Cedex 13, France.
} 
\email{huguet@apc.univ-paris7.fr, jacques.renaud@univ-mlv.fr}

\date{\today}


\begin{abstract}
We show that the Laplace-Beltrami equation $\square_6 a =j$ in $(\setR^6,\eta)$, $\eta := \mathrm{diag}(+----+)$, leads under very moderate assumptions to both 
the Maxwell equations and the conformal Eastwood-Singer gauge condition on conformally flat spaces including the spaces with a Robertson-Walker metric. 
This result is obtained through a geometric 
formalism which gives, as byproduct, simplified calculations. In particular, we build an atlas for all the conformally flat spaces considered which
allows us to fully exploit the Weyl rescalling to Minkowski space.
\end{abstract}

\maketitle

\section{Introduction}\label{Sec-Introduction}

This paper describes a geometrical framework for the study of conformally invariant fields in conformally 
flat spaces in four dimensions. Applications to scalar and electromagnetic 
fields are made. In particular, we show that the equation $\square_6 a =j$ in $(\setR^6,\eta)$, $\eta := \mathrm{diag}(+----+)$, 
where $\square_6$ is the Laplace-Beltrami operator, leads under very 
moderate assumptions to both the Maxwell equations and the conformal Eastwood-Singer gauge conditions \cite{EastwoodSinger} on $4$D conformally 
flat spaces. In addition, our proof of this result allows us to propose a new fiber bundle in which, broadly speaking,  the Maxwell equations on a 
conformally flat space  are converted into constrained scalar equations on Minkowski space.  This  drastically simplifies the practical calculations.  

The basic geometrical idea is to build four-dimensional spaces as the intersection in $(\setR^6,\eta)$, of a surface and the 
five-dimensional null cone (invariant under the linear conformal group SO$_0(2,4)$). The metric on such a space is induced 
from that of $(\setR^6,\eta)$. In particular, for any given Robertson-Walker (RW) metric, one can always find a surface such that the induced 
metric is the RW metric. Now, to each point of a space obtained in this way
corresponds a half line of the cone, hence all spaces can be realized as subsets of the set of the half lines: the cone modulo the 
dilations. In addition, in the intersection of two such subsets the spaces are related through a Weyl rescaling. 
As a special case, the Minkowski space can be obtained in that scheme. This allows us to build a particular atlas of the cone modulo the dilations.
In effect, copies of that Minkowski space can be obtained
by displacing the surface intersecting the cone thanks to the action of O$(2,4)$. This generates a covering of the set of the half lines. 
One can go a step further by introducing coordinates of $\setR^6$ with the specific property
that their restriction to a particular Minkowski space of the covering are the usual Minkowskian Cartesian coordinates. 
Each space of the covering together with its corresponding set of Minkowskian Cartesian coordinates can be turned into a
local chart of the cone modulo the dilations, which is thus endowed with a Minkowskian atlas. This atlas is one of 
the key ingredients to perform simple calculations. In particular, it allows us to use the local Weyl rescaling to handle 
global problems.   

The study of conformal fields, as the electromagnetic field, viewed as restrictions to four dimensional spaces of 
fields on $\setR^6$ can be traced back to the seminal paper of Dirac \cite{Dirac6cone}. There, he introduces the ``six cone formalism''
which sets the conditions, mainly the homogeneity of the fields on $\setR^6$, to obtain conformal Minkowskian fields. This approach
takes into account the SO$_0(2,4)$ symmetry of the Minkowskian equations under consideration from the beginning. This formalism has been extended 
in group theoretical context by Mack and Salam \cite{MS} almost $35$ years later and in study of conformal generalizations of QED on Minkowski and 
Anti-de~Sitter spaces in the mid 80's by many Authors (see \cite{Bayenetal} and references herein). 
The fact that the conformal group  SO$_0(2,4)$ includes as 
subgroups, besides the Poincar\'e and the anti-de~Sitter groups, the de~Sitter group was our starting point in the study of the relation between
conformal scalar field in the de~Sitter and Minkowski space \cite{pconf1},\cite{pconf2}. The generalization to the electromagnetic field
in a conformal gauge on the de~Sitter space was tackled in  \cite{pconf4}. There following \cite{Bayenetal} we used a method of auxiliary fields to obtain
a two-point function in a conformal gauge which reduces to the Eastwood-Singer gauge \cite{EastwoodSinger}.  These works was concerned by
free quantum fields, but part of the methods used also applied in the case of classical fields with sources. In that context, the tools 
previously build, especially the use of Minkowskian charts on the set of half lines, allowed us to reproduce, with simplified calculations, the result obtained by   
Higuchi and Cheong \cite{HiguchiCheong-1} for the problem of two charges in de~Sitter space. The geometrical framework we build up in the 
present work encompass and generalizes the method we developed in these previous studies. It extends the formalism to a class of conformally
flat spaces which contains in particular the spaces endowed with a RW metric. It provides an atlas which permits a global use of the Weyl relation 
to Minkowski space. Last but not least, it also provides a deeper view on the geometrical nature of the objects involved.

This article is organized as follows. The Sec. \ref{Sec-Geometry} is devoted to the geometrical framework. We first gives a global view
and motivate the main definitions. Then we turn to the geometrical formulation and consequences on tensors fields of the assumptions of
transversality and homogeneity. This allows us to see how the Weyl rescaling emerges in this context. The action of 
the SO$_0(2,4)$ group on tensor fields is then discussed. The construction of the Minkowskian atlas, and the expression of some 
properties in it follows.  This section ends by the proof that all the RW metrics can be obtained in the present formalism.
In Sec. \ref{SEC-ScalarField}, we discuss the conformal scalar fields, we show how both equations and
fields on $\setR^6$ and on a conformally flat space are related. The Sec. \ref{Sec-Maxwell} contains the proof of the proposition
that the Maxwell equations and the Eastwood-Singer gauge condition are obtained from homogeneous one form fields satisfying the 
Laplace-Beltrami in $\setR^6$  and some transversality requirement. Although the result is purely geometrical, the proof makes use 
of calculations performed in the Minkowskian atlas. This make apparent the practical calculation method inherited from this geometrical framework. Some 
generalizations are discussed in Sec. \ref{SEC-SomeCom}. 
Some properties of homogeneous tensors are reminded in Appendix \ref{Sec-HomogTensor}. Appendix \ref{Sec-AppCoordMink} resembles 
a few additional comments on some particular coordinate systems for the sake of completeness. Definitions and conventions are collected 
in Appendix \ref{Sec-AppendiceDef}. For convenience, the conventions for indices are repeated here:
\begin{eqnarray*}
\alpha, \beta, \gamma, \delta, \ldots &=&0, \ldots, 5,\\
\mu,\nu,\rho,\sigma\ldots &=&  0, \ldots, 3,\\
i, j, k, l, \ldots &=& 1, \ldots, 3,\\
I,J, \ldots&=& c, 0,\ldots,3,+.
\end{eqnarray*} 
The space $\setR^6$ is provided with the metric  $\eta~=~\mbox{diag}(+,-,-,-,-,+)$. A point of $\setR^6$ is denoted by $x$ or $y$,
the letter $y$ refers most often to the linear structure of $\setR^6$.

\section{Geometry}\label{Sec-Geometry}
\subsection{Overview of the formalism and definitions}\label{Sec-OverviewFormalism}

This section is intended to give the Reader a global view of the formalism. We motivate the introduction of key structures as the cone modulo 
the dilations or the bundle $B_f$. We give their definitions and their main properties, leaving the proofs to the forthcoming sections. 
For convenience a reminder of definitions and conventions is given in Appendix \ref{Sec-AppendiceDef}.
  
\subsubsection{The general framework}
\begin{figure}[ht]
 \begin{center}
 \includegraphics[width = 7.5cm]{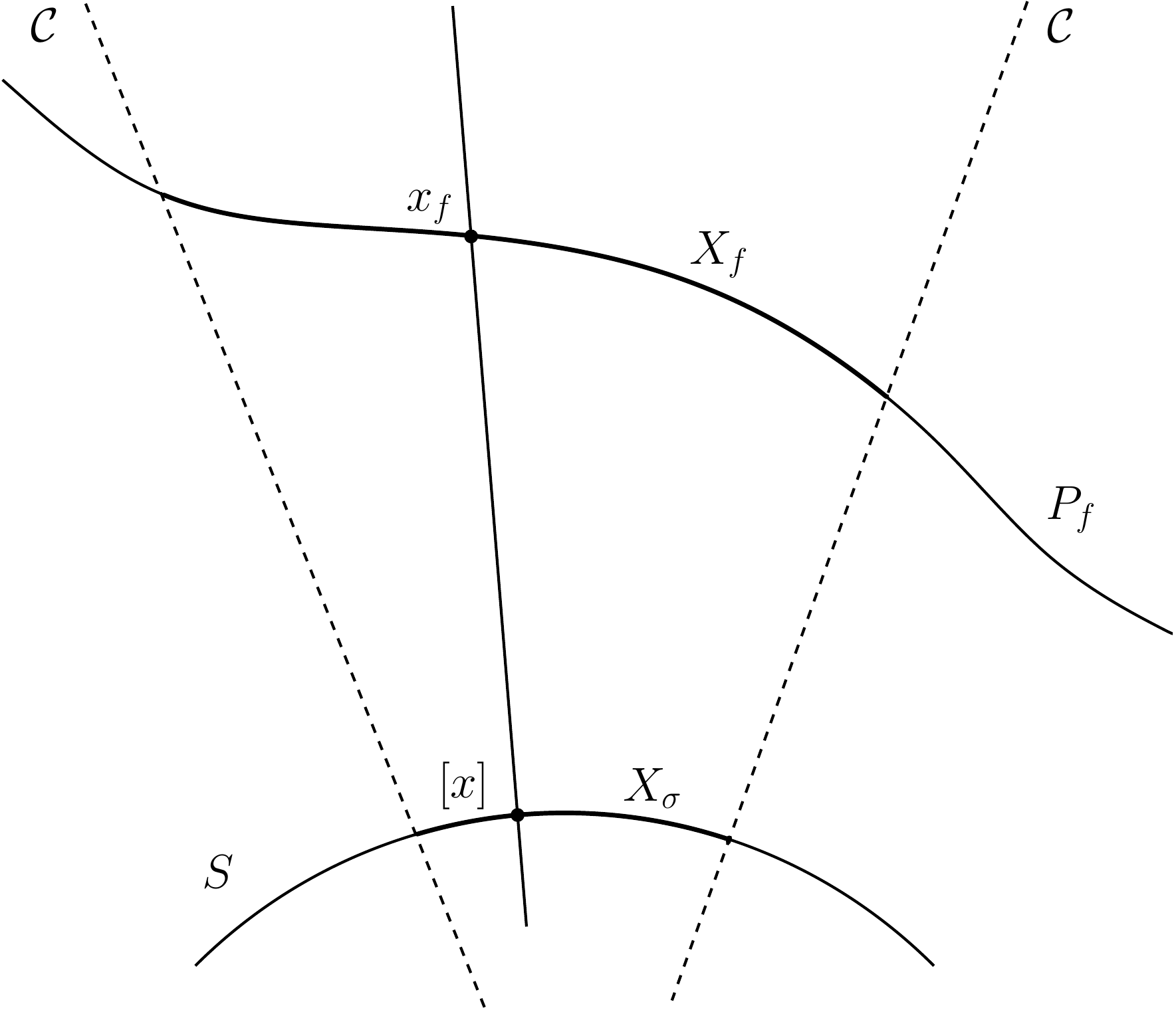}
 \caption{The main construction: A space $X_f$ (bold line) is obtained as the intersection $P_f\cap {\cal C}$ of $P_f$ a submanifold of $\setR^6$ defined by 
the equation $f(x) = 1$, $x\in\setR^6$ and the five-dimensional null cone ${\cal C}$. To each point of $X_f$  corresponds a 
 unique half-line $[x]$, that is a point of ${\cal C'}$ which can be identified with a point of the intersection (bold line) $X_\sigma$  
of ${\cal C}$ with the sphere $S$.}\label{Fig-GlobalPicture}
 \end{center}
\end{figure}

The main geometrical construction is pictured in Fig. \ref{Fig-GlobalPicture}.
A four dimensional  manifold $X_f$ is obtained as the intersection of the five 
dimensional null cone $\mathcal{C}$ of $\setR^6$ and a surface  $P_f:= \{y\in \setR^6 : f(y) = 1\}$, in which $f$ 
is homogeneous of degree one. 
On a manifold $X_f$ each point is intercepted by a
single half line of $\mathcal{C}$. Conversely, thanks to the homogeneity of $f$, each half line intercepts at most once the manifold $X_f$.
As a consequence each manifold $X_f$ can be realized as a 
subset $X'_f$ of a single set: the set $\mathcal{C'}$ of the half lines of $\mathcal{C}$, namely the cone modulo the dilations. 
This construction can be translated in relations between spaces which are collected in the following diagram
\begin{equation}\label{EQ-lfde[x]}
 \xymatrix{
\setR^6 &\\
X_f \ar[u]^{l_f} &X'_f\subset \mathcal{C'} \ar[ul]_{l'_f} \ar[l]^{\lambda_f}
}
\end{equation}
in which
\begin{align}\label{Eq-Deflpf}
l'_f: X'_f  &\rightarrow \setR^6\\
       [x] &\mapsto x_f := [x] \cap X_f, \nonumber
\end{align}
where $[x]$ is the half line of $\mathcal{C}$ (that is a point of $\mathcal{C'}$) which contains the point $x$,
\begin{align}\label{Eq-Defif}
l_f: X_f  &\rightarrow \setR^6\\
        x&\mapsto l_f(x) = x, \nonumber
\end{align}
is the canonical injection, and
\begin{align}\label{Eq-Deflamdaf}
\lambda_f: X'_f & \rightarrow X_f\\
           [x]&\mapsto x_f := [x] \cap X_f,\nonumber
\end{align}  
which is a diffeomorphism.
Note that, thanks to the homogeneity of $f$ one has $f(x/f(x))=1$, and thus
\begin{equation}\label{Eq-xf}
x_f=\frac{x}{f(x)}.
\end{equation}

Now, a set of  half lines may cross several manifolds $X_f$, in other words the realizations $X'_f$ on $\mathcal{C'}$ of these manifolds 
have a non-empty intersection, in that case  one can define some common coordinate systems (at
least locally). 
This property allows us to drastically simplify many practical calculations, this can be explained as follows. 
Firstly, as we will prove in  Sec. \ref{Sec-MinkAtlas}, the cone modulo the dilations as a manifold 
can be covered by a collection of realizations $X'_f$ of Minkowski spaces, each endowed with a
Cartesian system of coordinates. We will call  Minkowskian atlas this covering together with these coordinates, and Minkowskian charts 
the elements of this atlas. Secondly, we will prove in Sec. \ref{Sec-WeylRelation} that
in their intersection two realizations $X'_f$ are related through a Weyl rescaling. Finally,
due to the existence of the above covering all the manifolds $X_f$ are Weyl related to a Minkowski space, that is are 
conformally flat. In addition, and (this is the point for the simplifications) a conformal equation reads under its Minkowskian 
form in a Minkowskian chart.

Amongst the spaces obtained within this scheme the
space $X_\sigma$, which is obtained as the intersection of the cone with the $5$-sphere $S$,
provides a realization of the abstract manifold $\mathcal{C'}$ itself. Indeed,  $S$ is  intercepted only once by each half line of $\mathcal{C}$ 
(Fig. \ref{Fig-GlobalPicture}). We will use this 
realization of $\mathcal{C'}$, which is included in the Einstein space, for drawing conformal diagrams in Sec.
\ref{Sec-MinkAtlas} and some practical 
calculations (Sec. \ref{SEC-ScalarField}).

\subsubsection{Geometry and the Maxwell field}
Other important structures, more specifically related to the Maxwell field,  have to be introduced. 
In the scheme we use, which  is inspired by the Dirac's ``six cone formalism'' \cite{Dirac6cone}, the Maxwell field $A^f$, a one form 
field on $X_f$, is obtained from a one form field $a$ on $\setR^6$ homogeneous of degree zero. We remind (see Appendix \ref{Sec-HomogTensor})
that the Cartesian components of $a$ are then homogeneous of degree -1.  
In Sec.  \ref{Sec-Maxwell} we will prove the following property~: for $a$  homogeneous of degree zero, $j$ homogeneous of degree $-2$ satisfying 
\begin{equation}\label{EQ-Box6aETDieze-OverviewFormalism}
\left \{
\begin{aligned}
&\square_6 a = j\\
&\sharp_{\eta} a\vert_{\mathcal{C}} \in T(\mathcal{C})\\
&\sharp_{\eta} j\vert_{X_f} \in T(X_f),
\end{aligned}
\right.
\end{equation}
the field $A^f := l_f^*(a)$ and the current $J^f := l_f^*(j)$ defined on $X_f$ satisfy the Maxwell equations and the Eastwood-Singer gauge 
condition \cite{EastwoodSinger}~:
\begin{equation}\label{EQ-MaES-OverviewFormalism}
\left \{
\begin{aligned}
&\left(\square_f \delta^\nu_\mu - \nabla_\mu \nabla^\nu + R^\nu_{~~\mu}\right) A^f_\nu = J^f_\mu\\
&\left(\square_f\nabla^\nu - 2 \nabla_\mu \left(R^{\mu\nu} - \frac{1}{3}R g^{\mu\nu}\right)\right) A^f_\nu = 0,
\end{aligned}
\right.
\end{equation} 
where $\square_f$ is the Laplace-Beltrami operator on $X_f$.
Note that for maximally symmetric spaces  the Eastwood-Singer gauge condition reduces to the more familiar form
\begin{equation*}
 (\square_f + \frac{R}{6})\nabla \cdot A^f = 0,
\end{equation*}
with a constant Ricci scalar $R$. Note also that the two conditions appearing in (\ref{EQ-Box6aETDieze-OverviewFormalism}) mean that 
the vector fields $\sharp_{\eta} a = a^\alpha\partial_\alpha$ and $\sharp_{\eta} j = j^\alpha \partial_\alpha$ are tangent
respectively to $\mathcal{C}$ and to $X_f$. 

Beside the property by itself its proof makes use of a new geometrical structure: the fiber bundle
\begin{equation}\label{EQ-DefBf*}
 B_f^* := \bigcup_{x\in X_f} T_x^*(\setR^6).
\end{equation} It is  obtained in restricting the base space of the cotangent bundle $T^*(\setR^6)$ to $X_f$.
This object allows us to solve some problems 
in a more simple way than by tackling them directly on $X_f$.
It also describes more accurately the transition between 
the fields on $\setR^6$ and those on $X_f$. More precisely, although under the assumptions made in (\ref{EQ-Box6aETDieze-OverviewFormalism}) 
the Maxwell field $A^f$ is directly obtained as the pullback of $a$. The proof of this result involves an intermediate step in which 
the equations $\square_6 a = j$ are, in a sense which will be made precise in Sec. \ref{Sec-Maxwell}, restricted to  
the six conformal scalar equations
\begin{equation}\label{EQ-ScalConfaf-Overview}
 (\square^{(s)}_f + \frac{R}{6})a^f_\alpha = j^f_\alpha,
\end{equation}
where $\square^{(s)}_f$ is the scalar Laplace-Beltrami operator on $X_f$ and
in which both the $a^f$'s and the $j^f$'s belongs to the set $\mathcal{B}_f^*$ of the sections of  $B_f^*$. 
These equations reduce (when the two constraints of (\ref{EQ-Box6aETDieze-OverviewFormalism})
are applied) to the Maxwell equations and the Eastwood-Singer condition (\ref{EQ-MaES-OverviewFormalism}).
However they are more easier to handle. Even more, when quantizing the electromagnetic field, these equations account directly for the gauge fixing: 
part of the $\{a^f\}$'s components leads to the Maxwell field, the others carry constraints \cite{Binegardetal,Bayenetal,pconf3}.

The field $a^f$ is obtained from the field $a$ by restricting the base of $T^*(\setR^6)$ to $X_f$ and the field $A^f$ 
from the field $a^f$ by further restricting the fiber $T^*_x(\setR^6)$ to  $T^*_x(X_f)$.
The relations between $a, a^f$ and $A^f$ (also true for $j, j^f$ and $J^f$)  are collected in the following diagram
\begin{equation*}
\xymatrix{
a\in \Omega_1(\setR^6) \ar[d]_{\widetilde{l_f}} \ar[dr]^{l^*_f} & \\
a^f \in \mathcal{B}_f^* \ar[r]_{{r_f}} &\Omega_1(X_f) \ni A^f 
}
\end{equation*}
in which 
\begin{align}\label{Eq-Def-if-tilde}
\widetilde{l_f}~:  \Omega^1(\setR^6) &\rightarrow \mathcal{B}_f^*\\
                                   a &\mapsto a^f:=\widetilde{l_f}(a) \nonumber
\end{align}
such that:
$\forall\, x\in X_f, \forall\, V \in  T_x(\setR^6)$, $a^f(x) [V] = a(x) [V]$, 

\begin{align}\label{Eq-Def-rf-tilde}
{r_f}: \mathcal{B}^*_f &\rightarrow \Omega^1(X_f)\\
               a^f&\mapsto {r_f}(a^f)=A^f \nonumber
\end{align}
such that:
$\forall\, x\in X_f, \forall\, U \in  T_x(X_f)$, $A^f(x) [U] = a(x)[U] $.

\subsection{The transversality condition}\label{SEC-TransvCond}
This is a condition on tensor fields which will appear very often. 
Roughly speaking, for a tensor $t$ fulfilling this relation, the map $t\mapsto T^f$ has a good behavior with respect to group 
invariance, Weyl relation, equations... This condition reads as follows: A tensor
$t\in\mathcal{T}_p^0(\setR^6)$ fulfills the transversality condition if and only if, 
for any $x\in\cal C$ and for any $V_1,\ldots,V_p\in T^*_x({\cal C})$, $t(x)(V_1,\ldots,V_p)=0$\linebreak as soon as 
one of the arguments $V_i$ is equal to the dilation field $\xi := y^\alpha\partial_\alpha$, \ie~$V_i\propto \overrightarrow{Ox}$. 

Note that, when $t\in \Omega^p(\setR ^6)$, one often finds the little bit stronger condition $i_\xi t=0$  which,  for $p=1$, reduces  to
$y^\alpha t_\alpha=0$ (from which the name ``transversality condition'' originates after Dirac's paper \cite{Dirac6cone}, altough 
in the context of mechanics this kind of condition is usualy termed ``horizontality''), equivalently in
an index-free notation: $\sharp_{\eta} t\vert_{\mathcal{C}} \in T(\mathcal{C})$.
This condition will eventually be  required for any field considered here. For quantum fields, as usual when a gauge condition is present,  
the implementation of the constraint at the quantum level (in a Gupta-Bleuler quantization scheme) is done after the quantization.

\subsection{Homogeneity and Weyl relations}\label{Sec-WeylRelation}

In this section, we consider two manifolds, 
$X_f$ and $X_h$ defined, as in 
Sec. \ref{Sec-OverviewFormalism}, as the intersections of $\mathcal{C}$ and the manifolds $f(y) = 1,\mbox{ and } h(y)=1$.
Let $t\in\mathcal{T}_p^0(\setR^6)$ be an homogeneous tensor field of degree $d(t)$, 
we want to compare the fields $T^f:=l^*_f t$ and $T^h:=l^*_h t$. For this purpose we realize both  fields on $\mathcal{C'}$ and obtain a Weyl
relation between $T'^f:=l'^*_f t $ and $T'^h:=l'^*_h t $.
Let us show that if $t$ fulfills the transversality condition then, on $X'_f \cap X'_h$,
the following property holds:
\begin{equation}\label{EQ-WeylTenseursHomogenes}
\left(T'^f\right)\left([x]\right) = \left(K^f_h([x])\right)^{d(t)}  \left(T'^h\right)\left([x]\right).
\end{equation}
where we defined
\begin{equation}\label{EQ-defK}
 K^f_h([x]) := \frac{h(x)}{f(x)}. 
\end{equation}
Thus the function $K^f_h$ appears as the conformal factor. For further references we also define the related one form on $\mathcal{C'}$
\begin{equation}\label{EQ-defW}
 W^{(f,h)} := d \ln (K^f_h)^2.
\end{equation}
Note that, no reference to a metric structure is made in these two definitions.

In order to prove (\ref{EQ-WeylTenseursHomogenes}), we begin with showing that, for $x\in\mathcal{C}$ and $V' \in T_{[x]}(\mathcal{C'})$ one has
\begin{equation}\label{EQ-lf*etlh*}
f(x)l'_{f*}([x])\left[ V'\right] = h(x)l'_{h*}([x])\left[V'\right] + N,
\end{equation}
where $N$ is a null vector which belongs to a half-line $[x]$. We first remark that
\begin{equation*}
l'_h([x])= \frac{f(x)}{h(x)}l'_f([x])=K_f^h([x])l'_f([x]).
\end{equation*}
Then, differentiating the rightmost term, one obtains, for any $x\in\cal C$ and $V'\in T_{[x]}(\mathcal{C'})$,
\begin{equation*}
l'_{h*}([x])[V']= l'_{K_f^h*}([x])[V']l'_f([x])+K_f^h([x])l'_{f*}([x])[V'].
\end{equation*}
The first term of the r.h.s. of this expression is proportional to $x$, and as a consequence,  belongs to the half-line $[x]$. The expression 
(\ref{EQ-lf*etlh*}) follows at once.

Now, let us consider $V'_1,\ldots,V'_p \in T_{[x]}(\mathcal{C'})$. Then using in succession 
the definitions (\ref{Eq-Deflpf})-(\ref{Eq-Deflamdaf}), the homogeneity of $t$,
the property (\ref{EQ-lf*etlh*}) and finally the transversality condition  for $t$ (Sec. \ref{SEC-TransvCond}), one has
\begin{align*}
&T'^f([x])\left[V'_1,\ldots,V'_p\right]\\
&=\left(l'^*_f t\right)\left([x]\right)\left[V'_1,\ldots,V'_p\right]\\ 
&= t\left(\frac{x}{f(x)}\right)\left[l'_{f*}V'_1,\ldots,l'_{f*}V'_p\right]\\
&= \left(\frac{1}{f(x)}\right)^{d(t) - p}\!t(x)\left[l'_{f*}V'_1,\ldots,l'_{f*}V'_p\right]\\
&= \left(\frac{1}{f(x)}\right)^{d(t)}\!\!\!t(x)\left[f(x)l'_{f*}V'_1,\ldots,f(x)l'_{f*}V'_p\right]\\
&= \left(\frac{1}{f(x)}\right)^{d(t)}\!\!\!t(x)\left[h(x)l'_{h*}V'_1 + N_1,\ldots,h(x)l'_{h*}V'_p + N_p\right]\\
&=\left(\frac{1}{f(x)}\right)^{d(t)}\!\!\!t(x)\left[h(x)l'_{h*}V'_1,\ldots,h(x)l'_{h*}V'_p \right]\\
&=\left(\frac{h(x)}{f(x)}\right)^{d(t)}\!\!\!T'^h([x])\left[V'_1,\ldots,V'_p\right],\\
\end{align*}
which is the announced result.\pagebreak

Finally, thanks to the isomorphism $\lambda_f$ between the $X_f$'s and
the $X'_f$ and using the notation $T^f$ for the tensors on the $X'_f$'s instead of $T'^f$'s, one can recast 
the above relation  (\ref{EQ-WeylTenseursHomogenes}) 
under the more familiar form
\begin{equation}\label{EQ-WeylMetricXfXh-Short}
T^f(x) = \left(K^f_h(x)\right)^{d(t)}  T^h(x),
\end{equation}
in which $T^f(x)$ stands for $T'^f([x])=T^f(x_f)$. In particular, between metrics this relation specializes to
\begin{equation}\label{EQ-WeylMetricXfXh}
g^f(x) = \left(K^f_h(x)\right)^{2}  g^h(x),
\end{equation}
which makes apparent that the Eq. (\ref{EQ-WeylTenseursHomogenes}) is in fact a 
Weyl relation in the usual sense.

\subsection{Action of SO$_0(2,4)$}
In this section we specify the action of SO$_0(2,4)$ on the various objects defined above. In particular, we define the action of
SO$_0(2,4)$ on the sections $a^f$ of the fiber bundle $B^*_f$ and show that the transversality condition ensures the 
SO$_0(2,4)$ invariance of the construction.

The natural action $x\mapsto L_gx=g x$ of SO$_0(2,4)$ on $\setR^6$ yields an action $L_g^f$ on $X_f$ defined through:
\begin{equation*}
X_f\ni x\mapsto L_g^f x=(g x)/f(gx)\in X_f. 
\end{equation*}
Setting $\omega_g^f(x)=f(x)/f(gx)$ for any $x\in \setR^6$, one obtains the  
action of the conformal group on the space time:
\begin{equation}\label{EQ-actgroupXf}
X_f\ni x\mapsto L_g^fx=\omega_g^f(x)g x\in X_f.
\end{equation}

\subsubsection{Action of the group on the scalar fields}
Let us recall that, for a scalar field $\phi$ homogeneous of degree $r$ the operator $l^*_f$ is defined through:
\begin{equation}
l^*_f\phi(x)= \phi(l_f(x))=:\Phi^f(x).
\end{equation}
Also, the group acts on $\phi$ through the natural representation $L_g^*\phi(x)=\phi(g^{-1}x)$.
The field $\Phi^f$ is defined on $X_f$ and we now define the representation $(L^f)^c$
of the conformal group on it. We just impose that the operator $l^*_f$ intertwines $(L^f)^c$ and the natural representation
$L^{*}$ on $\phi$:\linebreak
\begin{equation}
(L_g^f)^c  l^*_f=     l^*_f(L_g)^*.
\end{equation}
That is to say, using that $f(g^{-1}x)=\omega_g^f(g^{-1}x)$ for any $x\in X_f$,
\begin{align*}
((L_g^f)^c\Phi^f)(x)&=(L_g^f)^c  l^*_f\phi(x)\\
  &=l^*_f(L_g)^*\phi(x)\\
  &=\phi(g^{-1}x)\\
  &=\phi\left(\frac{g^{-1}x}{f(g^{-1}x)}f(g^{-1}x)\right)\\
  &=\left(\omega_g^f(g^{-1} x)\right)^r\Phi^f(L^f_{g^{-1}}x),
\end{align*}
which is the well-known action of the conformal group on conformal pseudo scalar fields of weight $r$.

\subsubsection{Action of the group on the one-forms}

The group acts on a 1-form $a$ of $\setR^6$ through
\begin{equation}
(L_g)^*a(x)=g\cdot a(g^{-1}x),
\end{equation}
where $g\cdot a$ stands for the spinorial action on the 6 components of $a$.

We are going to define a representation $(L_g^f)^c$ of SO$_0(2,4)$ on the 1-forms of $X_f$ as well as a representation $\tilde L_g^f$ on the
 sections of $B_f^*$. 
Moreover we will prove that the map $r_f$ intertwines these representations as soon as the form $a$ fulfills the transversality condition:
\begin{equation}\label{EQ-intertwine}
(L_g^f)^cr^fa^f=r^f\tilde L_g^fa^f \mbox{ as soon as } i_\xi a=0,
\end{equation}
which reads
\begin{equation*}
\tilde L_g^fa^f(x)[U]= (L_g^f)^cA^f(x)[U],\forall x\in X_f,\ \forall U\in T_x(X_f),
\end{equation*}
as soon as $i_\xi a=0$.

We first build the representation  $\tilde L_g^f$ on the field $a^f$ in a very similar way as for the scalar field. We just 
impose that $\tilde l_f$ intertwines 
the representations $\tilde L_g^f$ and $(L_g)^*$
and obtain straightforwardly:
\begin{equation}\label{EQ-rho_tilde}
\tilde L_g^fa^f(x)[V]=(\omega_g^f(g^{-1}x))^{r-1}g\cdot a^f(L_{g^{-1}}^fx)[V],
\end{equation}
for any $V\in T_x(\setR^6)$, which is {\em a fortiori} true for any \linebreak$U\in T_x(X_f)$.

We now define the  representation $(L_g^f)^c$ of the group on the  forms $A^f$. 
As a result, we cannot define a representation on the $A^f$ in the same way as above. As we will see,
one must impose a condition on $a$. In place of this construction, we define directly the representation $(L_g^f)^c$ through
\begin{equation}
(L_g^f)^cA^f(x)[U]= (\omega_g^f(g^{-1}x))^rA^f(L_{g^-1}^fx)[(L_{g^{-1}}^f)_*U],
\end{equation}
for $x\in X_f$ and $U\in T_x(X_f)$.
 We begin with calculating $(L_g^f)_*$ on any $U\in T_x(X_f)$ using the Leibniz rule:
\begin{equation}
(L_g^f)_*(x)[U]=\frac{g\cdot U}{f(gx)}-g x\left(\frac{\partial f(g x)[U]}{f(gx)^2}\right).
\end{equation}
The crucial remark is that the second term of the r.h.s. belongs to a line of the cone. This term will be denoted as $F$ in the following. 
At this time we can set down the representation $(L_g^f)^c$, putting $\omega:=\omega_g^f(g^{-1}x)$ for the readability:
\begin{equation}\label{EQ-rho}
\begin{aligned}
(L_g^f)^cA^f(x)[U]&=\omega^rA^f(L^f_{g^{-1}}x)\left[\frac{1}{\omega} g^{-1}\cdot U+F\right]\\
&=\omega^ra(L^f_{g^{-1}}x)\left[\frac{1}{\omega} g^{-1}\cdot U+F\right]\\
&=\omega^ra(L^f_{g^{-1}}x)\left[\frac{1}{\omega} g^{-1}\cdot U\right]\\
&=(\omega_g^f(g^{-1}x))^{r-1}g\cdot a^f(L^f_{g^{-1}}x)[U],
\end{aligned}
\end{equation}
the before last equality being due to the transversality condition.
The result (\ref{EQ-intertwine}) follows immediately from (\ref{EQ-rho_tilde}) and (\ref{EQ-rho}).
As a consequence, the map $l_f^*$ intertwines the representations $(L_g^f)^c$ and $L_g^*$ as soon as $a$ fulfills the 
transversality condition.
Note that the identification  $X_f=X'_f$ allows to realize the group action on $X'_f$ as well.

\subsection{Minkowskian atlas and conformal diagrams}\label{Sec-MinkAtlas}

As explained in Sec. \ref{Sec-OverviewFormalism} all the spaces $X_f$ are realized as subsets $X'_f$ of $\mathcal{C'}$ through the 
one to one map $\lambda_f$. Then, one can use the Weyl relation (\ref{EQ-WeylMetricXfXh-Short}) in order to simplify practical calculations in problems involving
conformally invariant equations.  As well known, these calculations can be much simpler than those performed on $X_f$ especially when the second 
space $X_h$ is a Minkowskian one. Unfortunately, one cannot in general recover the space $X_f$ with only one Minkowskian space.
However,  we will prove that the covering of the whole $\mathcal{C'}$ 
is possible using four Minkowskian spaces. These can be endowed with coordinates 
systems in order to form an atlas. In addition, these systems can be chosen to be Cartesian Minkowskian coordinates  to 
makes calculations simpler.  Now, the formalism we develop use $\setR^6$ as framework, in practical calculations the coordinates basis 
on which fields can be expanded are chosen on  $\setR^6$. As a consequence, the Cartesian Minkowskian coordinates of the atlas 
have to be related to coordinates in $\setR^6$. In the sequel, we will call Minkowskian systems both the systems in  $\setR^6$ and those
deduced from them in the atlas of $\mathcal{C'}$. This atlas will also be named Minkowskian as well as charts that compose it. We will prove 
that one Minkowskian plane $P_{f_{\sss N}}$ together with one Minkowskian coordinate system, the $N$-system, are sufficient to 
generate the whole Minkowskian atlas. Note that, some general comments in relation with such kind of coordinates are made in Appendix 
\ref{Sec-AppCoordMink}.

First let us introduce a useful graphical representation of $\mathcal{C'}$ as conformal diagram. Remind that
$\mathcal{C'}$ can be realized as the intersection $X_\sigma$ of the $5$-sphere, obtained through 
$P_\sigma$, $\sigma:=\sqrt{\delta_{\alpha\beta}y^\alpha y^\beta/2}$, with the cone $\mathcal{C}$ (see Fig.~\ref{Fig-GlobalPicture}). Points
of $X_\sigma$ and $\mathcal{C'}$ are then identified through the map $\lambda_\sigma$. 
This space $X_\sigma$ is naturally endowed with a very convenient global coordinate system $\{\alpha, \beta, \theta, \varphi\}$ 
obtained by setting $r_1=r_2 = 1$ in the hyper-spherical coordinate system $\{r_1, r_2, \alpha, \beta, \theta, \varphi\}$ defined through
\begin{equation}\label{EQ-coordAlpBetsurC}
\left \{
 \begin{aligned}
  y^5 &= r_1 \cos \beta \\
  y^0 &= r_1 \sin \beta \\
  y^i &= r_2 \sin \alpha\; \omega^{i}(\theta,\varphi)\\
  y^4 &= r_2 \cos \alpha,
 \end{aligned}
\right .
\end{equation}
in which $\beta\in[-\pi,\pi[,\, \alpha,\theta\in[0,\pi], \varphi\in[0,2\pi]$ 
and $\omega^{i}(\theta, \varphi)$ correspond to the usual spherical coordinates on $S^2$. The condition $r_1 = r_2 =:r$ gives a system 
on the cone, $r=1$ gives a system on  $X_\sigma$. Now, since as a result of the considerations of the Sec. \ref{Sec-WeylRelation}
a space $X_f$ is related to $X_\sigma$ through a Weyl rescaling, the subsets $X'_f$ can be conformally pictured  
in the plane $(\alpha,\beta)$. In the sequel we will call ``$(\alpha,\beta)$-diagram'' this conformal mapping (see Fig. (2) for an example.). 
One may note that the metric element on $X_\sigma$ in the $\{\alpha, \beta, \theta, \varphi\}$ system is
\begin{equation}\label{EQ-MetricCsAlphBet}
ds^2 = d\beta^2 - d\alpha^2 - \sin^2\alpha d\omega^2. 
\end{equation} 
This form shows that $X_\sigma$ is indeed included ($\beta$ belongs to a subset of $\setR$) in the static Einstein space of positive curvature.

Now, let us consider the space $X_{f_{\sss N}}$, $f_{\sss N}:=(y^5+y^4)/2 = 1$. As shown in \cite{pconf1} this space is 
a Minkowski space. The coordinate system $\{x_{\sss N}^{I}\},~I=c,\mu,+$ defined by 
\begin{equation}\label{EQ-coordNord}
\left \{
 \begin{aligned}
  x_{\sss N}^c &= \frac{y^\alpha y_\alpha}{4 f_{\sss N}^2(y^4,y^5)}\\
  x_{\sss N}^\mu &=  \frac{y^\mu}{f_{\sss N}(y^4,y^5)}\\
  x_{\sss N}^+ &= f_{\sss N}(y^4,y^5),
 \end{aligned}
\right .
\end{equation} 
(whose inverse is given for convenience in Appendix \ref{Sec-AppCoordMink})
provides on $X'_{f_{\sss N}} \subset\mathcal{C'}$ a chart, called $N$-chart in the sequel.
The restriction to the cone is obtained for $x^c_{\sss N} = 0$ and 
that for $P_{f_{\sss N}}$ for $x^+_{\sss N} = 1$.  
The metric induced from $\setR^6$ 
on $X'_{f_{\sss N}}$ is, in the $\{x^\mu_{\sss N}\}$-coordinate basis,
 $\eta_{\mu\nu}$, the usual form of the Minkowski metric.

\begin{figure}[ht]
 \begin{center}
\includegraphics[width = 5.cm]{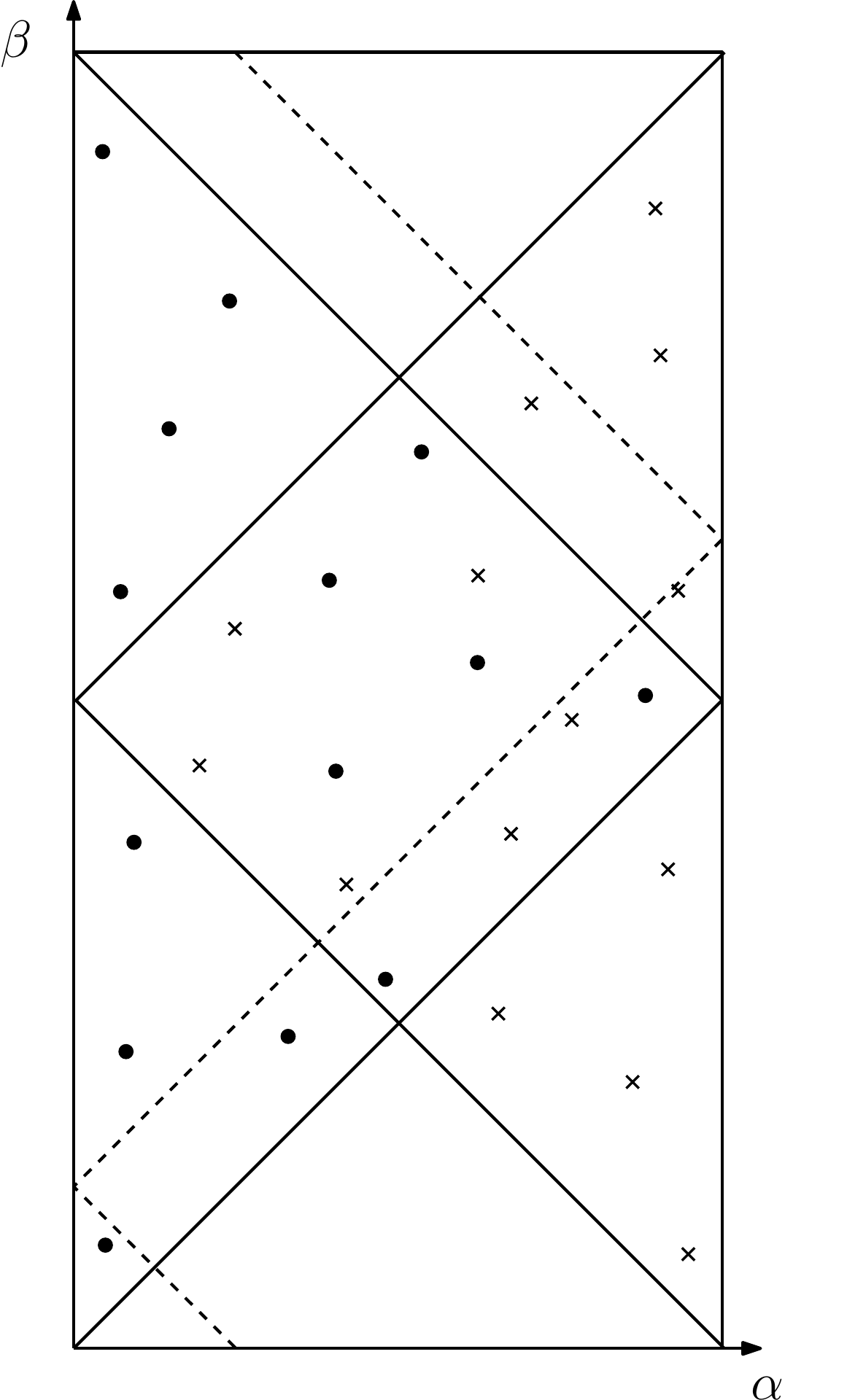} 
 \caption{The Minkowskian chart $N$ (dotted region) and $S$ (crossed region) obtained respectively from the Minkowski spaces 
$X_{f_{\sss N}}$, $f_{\sss N} := \frac{1}{2}(y^5+y^4)=1$ and $X_{f_{\sss S}}$, $f _{\sss S}:= \frac{1}{2}(y^5-y^4)=1$. 
These two charts can be translated along the the coordinate 
$\beta$ leading to a complete covering of $\mathcal{C'}$. This translation belongs to SO$_0(2,4)$ and corresponds to moving the planes 
$P_{f_{\sss N}}$,  $P_{f_{\sss S}}$ in $\setR^6$. The dashed lines are boundaries of a moved plane, the interior of that chart is 
on the l.h.s. (moved N-plane) or the r.h.s. (moved S-plane).}\label{Fig-NorthAndSouthChart}
 \end{center}
\end{figure}

The points where the above system becomes singular, that is on the subset $\{y^5+y^4 = 0\}$, 
correspond to points at infinity in the Minkowski space $X_{f_{\sss N}}$. They are mapped 
to the boundaries of the $N$-chart on the conformal $(\alpha, \beta)$-diagram in Fig. (\ref{Fig-NorthAndSouthChart}). 
Similar considerations apply to the space $X_{f_{\sss S}}$ defined by $f_{\sss S}:=(y^5-y^4)/2 = 1$ and the coordinate 
system $\{x_{\sss S}^{I}\},~I=c,\mu,+$ is obtained by replacing $N$ by $S$ in (\ref{EQ-coordNord}). 

A set of Minkowskian charts covering $\mathcal{C'}$ is finally obtained by moving the two surfaces $P_{f_{\sss N}}$, $P_{f_{\sss S}}$ with  
some elements of SO$_0(2,4)$. 
More precisely, we obtain new Minkowski spaces through the action of the one parameter subgroup of SO$_0(2,4)$ generated by \linebreak 
$X_{50} := y_5\partial_0 - y_0\partial_5=\partial_\beta$. The manifolds $X_{f_{\sss \beta N}}$ (resp. $X_{f_{\sss \beta S}}$) defined through
\begin{align*}
f_{\beta N}(y):=& f_{\sss N}(e^{-\beta X_{50}}y)=\frac{1}{2}\left(\cos\beta y^5 + \sin\beta y^0 + y^4\right) = 1,\\
f_{\beta S}(y):=& f_S(e^{-\beta X_{50}}y)=\frac{1}{2}\left(\cos\beta y^5 + \sin\beta y^0 - y^4\right) = 1.
\end{align*}
are Minkowski spaces. Corresponding to each case, one can define a coordinate system $\{x_{\sss\beta N}^{I}\}$ analogous to 
$\{x_{\sss N}^{I}\}$, replacing $f_{\sss N}$ by $f_{\sss \beta N}$ in (\ref{EQ-coordNord}) and similarly for $S$ in place of $N$. As a consequence, 
we obtain an atlas of $\mathcal{C'}$, in the usual sense, in which the systems of coordinates make the Weyl relations very transparent. 
Now every $X'_f$ space  is an open subset of $\mathcal{C'}$ and can be endowed with the above atlas.

As a final remark, the space $P_{f_{\sss S}}$ can be obtained from the space $P_{f_{\sss N}}$ through the transformation $y^4\mapsto -y^4$ 
which is also an element of O$(2,4)$. As a consequence, all the elements of our Minkowskian atlas can be obtained from $P_{f_{\sss N}}$ through 
a transformation of O$(2,4)$ that is to say through an (inoffensive) Cartesian isometric change of variables in $\setR^6$. 
Thus, the whole atlas is generated from $P_{f_{\sss N}}$ and the $N$-coordinates as announced in the beginning of this section. 

\subsubsection*{Some explicit conditions in the $N$-charts}\label{SEC-TransvCondCN}
We note that, 
 for $x\in X_f$,
\begin{equation}\label{EQ-x+f=K}
 x_{\sss N}^+ (x) =f_{\sss N}(x)=f_{\sss N}\left( \frac{x}{f(x)}\right)=K^f_{f_{\sss N}}(x).
\end{equation}

Also, in $N$-coordinates the dilation field reads \linebreak$\xi := y^\alpha\partial_\alpha = x_{\sss N}^+\partial_{x_{\sss N}^+}$. 

Now, a one-form $t\in \Omega^p(\setR ^6)$ homogeneous of degree zero, is expanded in the coordinate basis 
$\{dx_{\sss N}^c,dx_{\sss N}^\mu,dx_{\sss N}^+\}$  as 
\begin{equation*}
 t^{\sss N}= t^{\sss N}_c dx_{\sss N}^c + t^{\sss N}_\mu dx_{\sss N}^\mu + \frac{t^{\sss N}_+}{x_{\sss N}^+} dx_{\sss N}^+.
\end{equation*}
The last term is for convenience divided by $x_{\sss N}^+$ in order that $t^{\sss N}_+$ be of the same degree of homogeneity (zero) as the
other components. One has 
\begin{equation}  \label{EQ-TransvCondCN}
t^{\sss N}_+ = y^\alpha t_\alpha =i_\xi t.
\end{equation}  
In particular, the transversality condition  reads $t^{\sss N}_+ = 0$. The same result applies in others Minkowskian coordinates.

\subsection{Robertson-Walker metric}\label{SEC-RW}

Here we prove that for any Robertson-Walker (RW) metric $g^{\sss RW}$ one can find a function $f_{\sss RW}$ such that the metric induced 
from $\setR^6$ on the conformally flat space $X_{f_{\sss RW}}:= P_{f_{\sss RW}}\cap\mathcal{C}$ be the Robertson-Walker metric $g^{\sss RW}$.
In other words, each  Robertson Walker space can be realized at least locally as a space $X_{f_{\sss RW}}$.

Following Ibison \cite{Ibison} the set called $\{\mbox{conformal}(r,t)\}$ of conformally flat metrics defined through~:
\begin{equation}\label{EQ-rtMetric}
ds^2 = A^2(x^0,\norm{\boldsymbol{x}})dx^2;\;  dx^2 = \eta_{\mu\nu}dx^\mu dx^\nu,
\end{equation}
$A^2$ being a conformal factor and $\{x^\mu\}$ the Cartesian Minkowskian coordinates, includes the RW metrics.  
The notation $(x^0,\norm{\boldsymbol{x}})$ for the arguments of the function $A$ indicates that it depends 
separately of the time coordinates and of the space coordinates  and that the space coordinates appears only through the radius 
$\norm{\boldsymbol{x}}$.  

Let us assume now that $A$ for some RW metric is given in a Minkowskian coordinate system, say the system  $\{x_{\sss N}^{\sss I}\}$ (\ref{EQ-coordNord}).
The function
\begin{equation}
f_{\sss RW}(x) := \frac{x^+_{\sss N}}{A(x^0_{\sss N},\norm{\boldsymbol{x}_{\sss N}})},
\end{equation}
is invariant under the subgroup SO$_0(3)$ of SO$_0(2,4)$ generated by $y_i \partial_j - y_j \partial_i$. The equation $f_{\sss RW}(x) = 1$ defines a 
SO$_0(3)$ invariant space endowed with a metric which satisfies (\ref{EQ-rtMetric}), thus a RW space.

\section{Scalar fields}\label{SEC-ScalarField}
In this section, we consider scalar fields on $\setR^6$, homogeneous of degree $r$. These fields yield scalar fields on $X_f$. 
We  show that these fields fulfill the scalar conformal equation as 
soon as the original field fulfill the Laplace-Beltrami equation on $\setR^6$.

Let $\phi$ and $j$ two scalar fields on $\setR^6$ homogeneous of degree -1 and -3 respectively, and suppose that
\begin{equation}\label{Eq-EquationScalaireR6}
\square_6\phi=j.
\end{equation}
We claim that 
\begin{equation}\label{Eq-EquationScalaireXpf}
(\square_f+\frac{R}{6})\Phi^f=J^f,
\end{equation}
where $\square_f$ is the Laplace-Beltrami operator  and $R$ the scalar curvature for the space $X_f$.
 We consider $X_\sigma$ the realization of the cone up to the dilation 
defined Sec. \ref{Sec-MinkAtlas}, on which all $X_f$ spaces realize as subsets.
Setting $u=r_1/r_2$ and $v=1/r_2$ in the hyperspherical coordinates (\ref{EQ-coordAlpBetsurC}) and using 
the homogeneity of $\phi$, the equation (\ref{Eq-EquationScalaireR6}) reads
\begin{equation}\label{Eq-EquationScalaireCoordr1r2}
v^2\left((1-u^2)\partial^2_u+\frac{1}{u}(1-u^2)\partial_u+1+\frac{1}{u^2}\partial^2_\beta-\Delta_{S^3}\right)\phi=j.
\end{equation}
The restriction to $X_\sigma$ ($u=v=1$) then gives
\begin{equation}\label{Eq-EquationScalairespere}
(\partial^2_\beta-\Delta_{S^3}+1)\Phi^\sigma=J^\sigma,
\end{equation}
which is the desired result. We obtain the similar result for the other hyper surfaces $X_f$ by using the Weyl correspondence between the spaces $X_f$.
Note that although straightforward the change of variables leading to the equation (\ref{Eq-EquationScalaireCoordr1r2}) is rather cumbersome.

\section{The Maxwell field}\label{Sec-Maxwell}

In this section we deal with one-forms $a$ and $j$ of $\setR^6$ satisfying $\square_6 a=j$ and such that $a$ is homogeneous of degree 0
and $j$ homogeneous of degree $-2$ (remind that this implies that $a_\alpha$ and $j_\alpha$ are  homogeneous of degree $-1$ and $-3$ respectively). 
We first show how in the present formalism the equation $\square_6 a=j$ leads to a set of six copies of the  equation of the conformal scalar field.
Then, we give the proof that  $A^f\in T^*(X_f)$ satisfies the Maxwell 
equations as soon as $a$ fulfills the transversality condition, and that an additional condition on $j$ allows us to recover a conformal gauge condition.

\subsection{Equation on $\mathcal{B}^*_f$}
Suppose that $B_f^*$ is endowed with a coordinate system $(x^I, dy^\alpha)$ where $x^I$ is any coordinate system of $X_f$ 
and $dy^\alpha$ a Cartesian system of coordinates on $T_x^*(\setR^6)$. In such a coordinate system the equation on $a^f$ reads
\begin{equation}\label{EQ-equationsurfibre}
 (\square^{(s)}_f + \frac{R}{6})a^f_\alpha = j^f_\alpha,
\end{equation}
where $\square^{(s)}_f$ is the scalar Laplace-Beltrami on $X_f$ and $R$ the Ricci scalar. This is a straightforward consequence of the fact that, on
$\setR^6$ in Cartesian coordinates one has $(\square_6a)_\alpha = \square_6^{(s)}a_\alpha$: since 
in a Cartesian coordinate system $\{y^\alpha\}$ of $\setR^6$ the equation $\square_6a=j$  reduces to six 
copies (one per components of $a$ and $j$) of the scalar equation (\ref{Eq-EquationScalaireR6}), each components $a_\alpha$ and  $j_\alpha$ 
being homogeneous of degree $d(a_\alpha) = -1$ and $d(j_\alpha) = -3$, one can thus apply the result (\ref{Eq-EquationScalaireXpf}) of the 
Sec. \ref{SEC-ScalarField}.

Two comments are in order concerning equations
(\ref{EQ-equationsurfibre}). In first, they do not present any gauge ambiguity and we proved in \cite{pconf3} that they 
allow a quantization of the Maxwell field on de Sitter space using a Gupta-Bleuler scheme where  
the condition of transversality is translated to a condition on states after quantization. 
In second, they can be used to solve  classical propagation problems for the Maxwell field: For instance, we considered in \cite{pconf4} 
the two-charges problem for the de Sitter space.

\subsection{Maxwell equation on $X_f$}
In this section, we implicitly identify the spaces $X_f$ and $X'_f$, writing all the objects on the common manifold ${\cal C}'$.

We now prove the statement of Sec. \ref{Sec-OverviewFormalism}: 
for $a$  homogeneous of degree zero, $j$ homogeneous of degree $-2$ satisfying 
\begin{equation}\label{EQ-Box6aETDieze-Maxwell}
\left \{
\begin{aligned}
&\square_6 a = j\\
&\sharp_{\eta} a\vert_{\mathcal{C}} \in T(\mathcal{C})\\
&\sharp_{\eta} j\vert_{X_f} \in T(X_f)
\end{aligned}
\right.
\end{equation}
the field $A^f := l_f^*(a)$ and the current $J^f = l_f^*(j)$ defined on $X_f$ satisfy the Maxwell equations and the Eastwood-Singer gauge 
condition \cite{EastwoodSinger}~:
\begin{equation}\label{EQ-MaES-Maxwell}
\left \{
\begin{aligned}
&\left(\square_f \delta^\nu_\mu - \nabla_\mu \nabla^\nu + R^\nu_{~~\mu}\right) A^f_\nu = J^f_\mu\\
&\left(\square_f\nabla^\nu - 2 \nabla_\mu \left(R^{\mu\nu} - \frac{1}{3}R g^{\mu\nu}\right)\right) A^f_\nu = 0.
\end{aligned}
\right.
\end{equation} 
We first prove the result on the  Minkowskian chart $X_{f_{\sss N}}$, after what, using the O$(2,4)$ invariance of the hypothesis and of the conclusion, and the 
properties of the Minkowskian charts,  we obtain this result on all the Minkowskian charts.
In the following, for readability, we note $a^{\sss N}, A^{\sss N}$, for $a^{f_N}$ and $A^{f_{\sss N}}$. We first consider the equation $\square_6 a = j$, 
and apply (\ref{EQ-equationsurfibre}) 
 on the Minkowski space $X_{f_{\sss N}}$ defined by $f_{\sss N}$ (used to build the N-chart),
this yields the system
\begin{equation}\label{Eq-EquationScalaire-aN}
\square_{\sss N}^{(s)}a^{\sss N}_\alpha =j^{\sss N}_\alpha,
\end{equation}     
where  $\square_{\sss N}^{(s)}$ is the scalar Laplace-Beltrami operator on $X_{f_{\sss N}} \simeq X'_{f_{\sss N}}$, $a^{\sss N}:=\widetilde{l_{\sss N}}(a)$ and 
$j^{\sss N}:=\widetilde{l_{\sss N}}(j)$.
As a second step we express in the $N$-chart of $\mathcal{C'}$ (\ref{EQ-coordNord}) the above system of equations (\ref{Eq-EquationScalaire-aN}). This
leads to the system already obtained in \cite{Binegardetal} (with slightly different notations)
\begin{equation}\label{EQ-SystAuxMink}
\left \{
\begin{aligned}
&\partial^2 a^{\sss N}_\mu + \partial_\mu   a_c^{\sss N}  = j_{\mu}^{\sss N} \\
&\partial^2  a_+^{\sss N}  - 2 \partial\cdot a^{\sss N} - 2 a_c^{\sss N} =  j_+^{\sss N}\\
&\partial^2 a^{\sss N}_c = j_c^{\sss N}.
\end{aligned}
\right .
\end{equation}

Then, we express the two constraints appearing in (\ref{EQ-Box6aETDieze-Maxwell}) in the $N$-chart. The first one is the transversality condition
applied to the field $a$, using the formula of ( \ref{EQ-TransvCondCN}) one has 
\begin{equation}\label{EQ-aplus=0}
 a_+^{\sss N} = 0.
\end{equation}
The second condition $\sharp_{\eta} j\vert_{X_f} \in T(X_f)$ implies the transversality of $j$ since  $T(X_f)\subset T(\mathcal{C})$, thus
\begin{equation}\label{EQ-jplus=0}
 j_+^{\sss N} = 0.
\end{equation}
In addition, $\sharp_{\eta} j\vert_{X_f} \in T(X_f)$ rewrites $ \sharp_{\eta} j [f] = 0$, which in the $N$-coordinates on the cone ($x_{\sss N}^c = 0$) reads
\begin{equation*}
  j^{\sss N}_c \dron{}{f}{x_{\sss N}^+}   + \frac{2~}{x_{\sss N}^+} \eta^{\mu\nu} j^{\sss N}_\mu \dron{}{f}{x_{\sss N}^\nu} = 0.
\end{equation*}
Thanks to the homogeneity of $f$ this equation rewrites
\begin{equation*}
  j^{\sss N}_c + 2~\eta^{\mu\nu} j^{\sss N}_\mu \frac{\partial_\nu f}{f}   = 0.
\end{equation*}
Then, using 
the definition (\ref{EQ-defW}), it becomes 
\begin{equation}\label{EQ-jcW}
j^{\sss N}_c = j^{\sss N} \cdot  W^{(f,f_{\sss N})},
\end{equation}
where the dot refer to the Minkowskian metric $\eta^f$.

Using the constraint (\ref{EQ-aplus=0}-\ref{EQ-jcW}), the system (\ref{EQ-SystAuxMink}) becomes
\begin{equation}\label{EQ-SystAuxMink-a+j+jc}
\left \{
\begin{aligned}
&\partial^2 a^{\sss N}_\mu - \partial_\mu  \partial\cdot a^{\sss N}   = j_{\mu}^{\sss N} \\
&\partial^2 \partial \cdot a^{\sss N} = j^{\sss N} \cdot W^{(f,f_{\sss N})}\\
& a_c^{\sss N} =  \partial \cdot a^{\sss N}.
\end{aligned}
\right .
\end{equation}
Owning to the map ${r_{f_{\sss N}}}$ the above system leads to
\begin{equation}\label{EQ-SystMink-W}
\left \{
\begin{aligned}
&\partial^2 A^{\sss N}_\mu - \partial_\mu  \partial\cdot A^{\sss N}   = J_{\mu}^{\sss N} \\
&\partial^2  \partial\cdot A^{\sss N} = J^{\sss N} \cdot W^{(f,f_{\sss N})}
\end{aligned}
\right .
\end{equation}
on the Minkowski space $X_{f_{\sss N}}$. Note that the map ${r_{f_{\sss N}}}$ becomes obvious 
in the N-coordinates (\ref{EQ-coordNord}):
for all $x \in X_f$ one has $A_\mu^f(x) = a_\mu^f(x) = a_\mu(x)$.

For the sake of argument, let us introduce a
self explanatory symbolical notation for the two operators appearing in the l.h.s. of
the Maxwell equations and the Eastwood-Singer gauge condition. With them the above system reads:
\begin{equation*}
\left \{
\begin{aligned}
& \left(M_{\sss N} [A^{\sss N}]\right)_\mu = J^{\sss N}_\mu\\
& ES_{\sss N}[A^{\sss N}] = J^{\sss N} \cdot W^{(f,f_{\sss N})},
\end{aligned}
\right .
\end{equation*} 
This result is available for the $N$-chart. Nevertheless, the O$(2,4)$ invariance of hypothesis and conclusion and the fact that
any Minkowskian chart can be deduced from the $N$-chart by mean of an O$(2,4)$ transformation, proves that this result is true for any 
Minkowskian chart of our atlas.

Now, using the same  symbolical notation
for the two operators appearing in the l.h.s of (\ref{EQ-MaES-Maxwell}), 
and following \cite{EastwoodSinger} to apply a usual (local) Weyl transformation between $X_{f_{\sss N}}$ and $X_f$ to each Eqs. of the 
system (\ref{EQ-SystMink-W}), one obtains 
\begin{equation*}
\left \{
\begin{aligned}
& \left(M_f [A^f]\right)_\mu = J^f_\mu\\
& ES_f[A^f] + W^{(f,f_{\sss N})} \cdot \left(M_f [A^f]\right)= J^{\sss f} \cdot W^{(f,f_{\sss N})},
\end{aligned}
\right .
\end{equation*} 
where the dots now refer to the metric $g^f$ and where we have taken into account the homogeneity of both the electromagnetic 
and the current one-form fields. They have respectively 
a degree of $-1$ and of $-3$ which correspond to conformal weights of zero and $-2$ for the components of the fields~: 
$A^f_\mu = A^{\sss N}_\mu$, $J^f_\mu = (K^f_{f_{\sss N}})^{-2} J^{\sss N}_\mu$. Finally, the second equation simplifies and we obtain the announced 
result on the $N$-chart. Then, using the above remark on O$(2,4)$ invariance, we obtain the result on the whole space.

Note that the Maxwell equations are obtained independently of the condition $\sharp_{\eta} j\vert_{X_f} \in T(X_f)$, which is used only 
to obtain the Eastwood-Singer gauge condition.  Note also that, in absence of source ($j=0$), the Eastwood-Singer 
gauge condition is
automatically fulfilled as soon as the transversality condition on $a$ is fulfilled.

It is important to point out that from a physical perspective the initial conditions and currents are given in the space $X_f$. In particular, the Minkowskian 
currents appearing in (\ref{EQ-SystMink-W}) are defined through the equation $J^f_\mu = (K^f_{f_{\sss N}})^{-2} J^{\sss N}$ in order
to satisfy the Weyl relation. They have in general no physical meaning in the Minkowski space. 

\section{Comments on some applications}\label{SEC-SomeCom}
Besides the results presented in the previous sections, the formalism depicted in the present paper explicits the geometrical 
nature of the various objects used in our previous works. 
This allows us to consider some straightforward generalizations, let us comment briefly about the classical and quantum situations.  

The method used for the classical propagation problem considered in \cite{pconf4}
extends naturally to the conformally flat spaces $X_f$. To summarize, given a set of initial conditions an currents in $X_f$ the problem 
of finding the $A^f$ solution of the Maxwell equations in the Eastwood-Singer gauge (or in a gauge contained in it) in $X_f$ amounts 
to use in the Minkowskian atlas the propagation formula established in \cite{pconf4},  which uses only the Minkowskian scalar Green's 
function. 

The quantization scheme used in \cite{pconf3} for the free electromagnetic field on de~Sitter space can be transposed
here for the most part. The definition of an SO$_0(2,4)$-invariant
scalar product on the space of the solutions of (\ref{EQ-equationsurfibre}) and the obtention (from the known mode solutions of  
(\ref{EQ-equationsurfibre})) of a reproducing kernel for the $a^f_\alpha$'s with respect to that product can be reproduced 
almost verbatim. The result is:
$\mathcal{W}^f_{\alpha\beta}(x,x')=-\eta_{\alpha\beta}D^+_f(x,x')$, where $D^+_f(x,x')$ is the scalar two-point function on $X_f$.
The construction of the Fock space, including the determination of the physical subspace is formally identical.   
The general form of a covariant two-point function, a bi-tensor on a specific $X_f$, 
requires more developments (note that, such a function  has been proposed recently in Minkowskian coordinates \cite{Faci}). 
This could be the object of future investigations.

\section*{Acknowledgements}
The authors thank C. Cachot for valuables discussions and  suggestions about the Secs.(\ref{Sec-MinkAtlas}) and (\ref{SEC-RW}). 

\appendix

\section{Homogeneous tensors}\label{Sec-HomogTensor}
For reference, we recall here some properties of an homogeneous $\left(^0_p\right)$-tensors $T$ of $\setR^6$. 

The set $\setR^6$, together with the scalar product $<y_1,y_2>$ $:= \eta_{\alpha\beta}y^\alpha y^\beta$, is naturally
endowed with a structure of linear space. A point in $\setR^6$ is located by a vector $y$ whose components in 
Cartesian coordinates are $\{y^\alpha\}$.  The linear structure allows us to identify the tangent spaces at any point through $\partial_\alpha (y)=
\partial_\alpha(0)$. This makes sense for expressions as $T(y)=T(\lambda y)$ since the tangent space at $y$ is identified with the tangent space at $\lambda y$.
Let us consider the dilatation map in $\setR^6$
\begin{align*}\label{Eq-DefRhoLambda}
\rho_\lambda: \setR^6 & \rightarrow \setR^6\\
           y&\mapsto \lambda y,
\end{align*}  
$\lambda$ being a positive real number. 
Since $T_y(\setR^6) \equiv \setR^6$ this is also the result of a push-forward on the vector $y$. In other
words $\rho_{\lambda *} \partial_\alpha = \lambda \partial_\alpha$.
For a $\left(^0_p\right)$-tensor field $T$ of the pullback reads $(\rho_{\lambda}^* T)(y) = \lambda^p T(\lambda y)$.
Now, an homogeneous $\left(^0_p\right)$-tensor field $T$ of degree $d(T)=r$  is 
defined through the relation
\begin{equation*}
(\rho_{\lambda}^* T)(y) =  \lambda^r T(y).
\end{equation*}
Thus,  homogeneous $\left(^0_p\right)$-tensor field $T$ of degree $r$ satisfy
\begin{equation}\label{EQ-TensHomoLienDefPullBack}
T(\lambda y) =  \lambda^{(r - p)} T(y).
\end{equation}

The homogeneity of a tensor field can be related to that of its components in some coordinate basis 
whose coordinates $\{x^{\sss I}(y)\}$ $I=0,\ldots,5$ are each one homogeneous functions of degree $d(I)$. Precisely, 
let us show that: for $T\in\mathcal{T}_p^0(\setR^6)$ a $\left(^0_p\right)$-tensor field of $\setR^6$, homogeneous of degree $d(T)$ one 
has
\begin{equation*}
d\left(T_{\sss I_1,\ldots,I_p}\right) =  d(T) - \sum_{k=1}^{p} d(I_k),
\end{equation*}
in particular for $x(y)=y$
\begin{equation*}
d\left(T_{\alpha_1,\ldots,\alpha_p}\right) =  d(T) - p.
\end{equation*}
This result is obvious in the particular case $x(y)=y$ since 
the Eq. (\ref{EQ-TensHomoLienDefPullBack}) reads in components
\begin{equation*}
T(\lambda y)_{\alpha_1,\ldots,\alpha_p}  =  \lambda^{(r - p)} T(y)_{\alpha_1,\ldots,\alpha_p}.
\end{equation*}
Now, moving to the more general coordinate basis $\{x^{\sss I}(y)\}$ on has
\begin{equation*}
T_{\sss I_1,\ldots,I_p}(\lambda y) = J^{\alpha_1}_{\sss I_1}(\lambda y) \ldots J^{\alpha_p}_{\sss I_p}(\lambda y)
T_{\alpha_1,\ldots,\alpha_p}(\lambda y)
\end{equation*}
where ${\displaystyle J^{\alpha}_{\sss I}(u) := \left(\dron{}{y^{\alpha}}{x^{\sss I}}\right)_u, u\in\setR^6} $,  
is homogeneous of degree $1 - d(I)$. Taking homogeneity into account in the r.h.s. of the above expression  leads to
\begin{equation*}
\begin{split}
T_{\sss I_1,\ldots,I_p}(\lambda y) =& \lambda^{\left(\sum_k(1-d(I_k)) + (d(T) - p)\right)} \\ 
\times& J^{\alpha_1}_{\sss I_1}(y) \ldots J^{\alpha_p}_{\sss I_p}(y)  T_{\alpha_1,\ldots,\alpha_p}(y),
\end{split}
\end{equation*}
from which the result follows.
\bigskip

\section{Notes on Minkowskian systems}\label{Sec-AppCoordMink}

The system appearing in Sec. \ref{Sec-MinkAtlas} has already been used in the literature (see for instance \cite{Binegardetal}),
it is reminiscent of systems called polyspherical systems (see for instance  \cite{kastrup}). One may note that they are framed 
in such a way that the ``extremal''coordinates $\{x^c,x^+\}$ are functions of the constraints (defining the cone $\mathcal{C}$ and 
the space $P_{f_{\sss N}}$ respectively) whereas the ``central'' coordinates  $\{x^\mu\}$ are those on the background space $X_f$. 
Cartesian Minkowskian coordinates can be obtained on $X_f$ setting $x^+:=f$ instead of $x^+:=f_{\sss N}$, with this choice
conformally invariant equations will appears on their Minkowskian form on $X_f$.  

For the de~Sitter space, one may verify that the $N$ and $S$ Minkowski charts correspond
to the stereographic projections from the North and South Poles from which their names originates.      

Finally,  we note for convenience that the inverse of system (\ref{EQ-coordNord}) reads
\begin{equation*}
\left \{
 \begin{aligned}
  y^5 &= x_{\sss N}^+ (1 + x_{\sss N}^c -\frac{1}{4} \eta_{\mu\nu}x_{\sss N}^\mu x_{\sss N}^\nu)\\
  y^4 &= x_{\sss N}^+ (1 - x_{\sss N}^c +\frac{1}{4} \eta_{\mu\nu}x_{\sss N}^\mu x_{\sss N}^\nu)\\
  y^\mu &= x_{\sss N}^+ x_{\sss N}^\mu.
 \end{aligned}
\right .
\end{equation*} 

\section{Conventions and definitions}\label{Sec-AppendiceDef}
We summarize here the conventions,  main structures and maps used in this paper.
They are discussed in Sec. \ref{Sec-OverviewFormalism}. 
Here are the conventions for indices:
\begin{eqnarray*}
\alpha, \beta, \gamma, \delta, \ldots &=&0, \ldots, 5,\\
\mu,\nu,\rho,\sigma\ldots &=&  0, \ldots, 3,\\
i, j, k, l, \ldots &=& 1, \ldots, 3,\\
I,J, \ldots&=& c, 0,\ldots,3,+.
\end{eqnarray*} 
The coefficients of the metric $\mbox{diag}(+,-,-,-,-,+)$ of $\setR^{6}$ are denoted $\eta_{\alpha \beta}$.
The definitions of spaces and maps reads~:
\begin{description}
\item $P_f := \{y\in \setR^6 : f(y) = 1\}$, in which $f$ is homogeneous of degree $1$.
\item $\mathcal{C} := \{y\in \setR^6 : (y^0)^2 - \boldsymbol{y}^2 - (y^4)^2 + (y^5)^2 = 0\}$, the null cone of $\setR^6$, 
\item $\mathcal{C'}$ denotes the set of the half lines of the cone $\mathcal{C}$, namely the cone modulo the dilations. One has
$[y] \in  \mathcal{C'}$ iff $[y]=\{u \in \mathcal{C} : \exists \lambda > 0, u = \lambda y\}$ 
\item $X_f = P_f \cap \mathcal{C}$, the physical space.
\item $X'_f = \{[x]\in\mathcal{C'} : \exists\, x \in X_f\}$ the realization of $X_f$ in $\mathcal{C'}$. 
\item $\displaystyle{B_f^* = \bigcup_{x\in X_f} T_x^*(\setR^6)}$
\item $\mathcal{B}_f^*$, the set of sections of  $B_f^*$
\item $l'_f: X'_f  \rightarrow \setR^6$\\
               $[x] \mapsto x_f := [x] \cap X_f$
\item $\lambda_f:X'_f  \rightarrow X_f$\\
               $[x]\mapsto x_f := [x] \cap X_f$
\item $l_f: X_f  \rightarrow \setR^6$\\
               $~x\mapsto l_f(x) = x$
\item $\widetilde{l_f}~:  \Omega_1(\setR^6) \rightarrow \mathcal{B}_f^*$\\
                $~~~~~~~~a \mapsto a^f:=\widetilde{l_f}(a)$\\
such that:\\
$\forall\, x\in X_f, \forall\, V \in  T_x(\setR^6)$, $a^f(x) [V] = a(x)[V]$. 
\item $r_f: \mathcal{B}^*_f \rightarrow  \Omega_1(X_f)$\\
               $~~a^f\mapsto {r_f}(a^f)=A^f$\\
such that:\\
 $\forall\, x\in X_f, \forall\, U \in  T_x(X_f)$, $A^f(x) [U] = a(x)[U]$, where $A^f:=l^*_f(a)$ is the physical 
field on $X_f$.
\end{description}

The Laplace-Beltrami operators on $(\setR^6,\eta)$ and $X_f$ are denoted respectively by $\square_6$ and $\square_f$. The operator $\square_f^{(s)}$ 
is the scalar Laplace-Beltrami operator on $X_f$: it acts on its argument as if it were a scalar.



\end{document}